%% file: main.tex
\documentclass[10pt,twocolumn,letterpaper]{article}

\usepackage[pagenumbers]{wacv} 				% To force page numbers, e.g. for an arXiv version

\input{preamble}

\definecolor{wacvblue}{rgb}{0.21,0.49,0.74}
\usepackage[pagebackref,breaklinks,colorlinks,allcolors=wacvblue]{hyperref}

\def\wacvPaperID{} % *** Enter the WACV Paper ID here
\def\confName{WACV}
\def\confYear{2027}

\title{A Self-Supervised Learning Framework for Video Encoding Complexity Clustering} 

\author{
	Krishna Srikar Durbha\\
	The University of Texas at Austin\\
	\and
	Hassene Tmar\\
	Meta Platforms, Inc.\\
	\and
	Ping-Hao Wu\\
	Meta Platforms, Inc.\\
	\and
	Ioannis Katsavounidis\\
	Meta Platforms, Inc.\\
	\and
	Alan C. Bovik\\
	University of Colorado at Boulder
}

\begin{document}

% Title
\maketitle

% Abstract
\begin{abstract}
	% Only 150 Words
	Adaptive video streaming is a widely used technique for delivering video content over the internet. One of the key challenges is determining the optimal encoding settings for each video, which can vary significantly based on its content and characteristics. In this paper, we propose Compression Echo Contrastive Learning (CECL), a novel self-supervised learning framework for clustering videos based on their encoding complexity. Our method leverages the response of a video to compression --- the `Compression Echo' --- as a supervisory signal, allowing the model to capture underlying encoding characteristics during pretraining. We conduct extensive experiments to demonstrate the effectiveness of our learned representations for the downstream task of clustering videos by their encoding complexity. Our results show that CECL improves upon existing state-of-the-art visual encoders and delivers strong bitrate and quality savings against the fixed bitrate ladder.\footnote{\scriptsize{The code will be made publicly available after the publication of the paper.}}
\end{abstract}

% Sections
\input{sections/1_introduction.tex}
\input{sections/2_related_works.tex}
\input{sections/3_motivation.tex}
\input{sections/4_methodology.tex}
\input{sections/5_experimental_setup_datasets.tex}
\input{sections/6_experiments_results.tex}
\input{sections/7_conclusion.tex}

\section*{Acknowledgements}
The authors thank the Texas Advanced Computing Center (TACC) at The University of Texas at Austin for providing HPC resources that have contributed to the research results reported in this paper. URL: http://www.tacc.utexas.edu.

% Appendix
% \clearpage
\appendix
\input{sections/appendix.tex}

% Bibliography
{
    \small
    \bibliographystyle{ieeenat_fullname}
    \bibliography{main}
}

\end{document}

%% file: preamble.tex
\usepackage{graphicx}
\usepackage{booktabs}
\usepackage{caption}
\usepackage{subcaption}
\usepackage{colortbl}
\usepackage{algorithm}
\usepackage{algpseudocode}
\usepackage{amsmath}

%% file: sections/1_introduction.tex
\section{Introduction and Motivation}
\label{sec:introduction}

\begin{figure*}[!ht]
	\centering
	\includegraphics[width=0.9\textwidth]{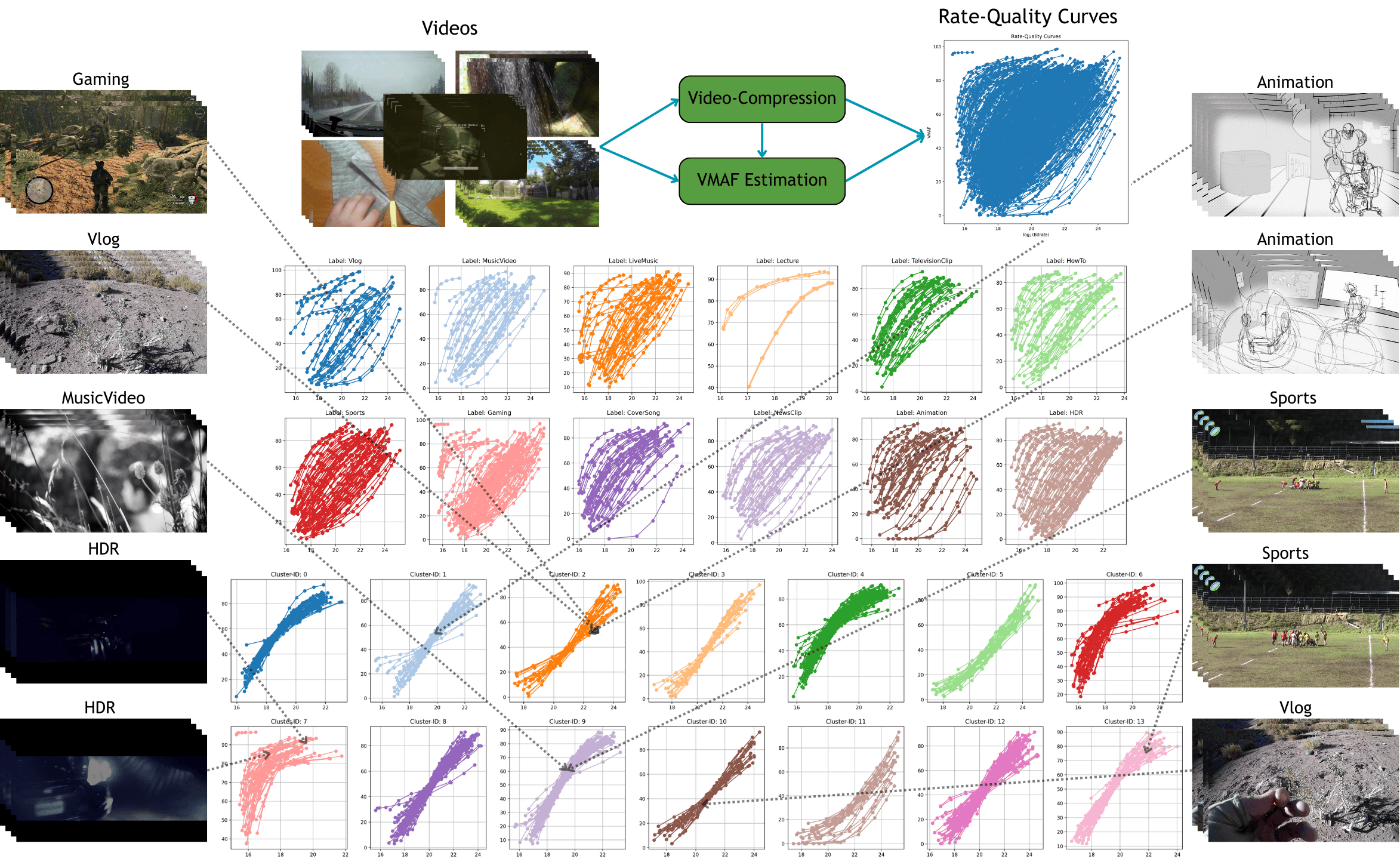}
	\caption{Rate-quality curves of the YouTube-UGC dataset grouped by semantic category labels (top) versus cluster labels obtained from Meta's clustering algorithm (bottom).}
	\label{fig:semantics_vs_encoding_complexity}
\end{figure*}

A recent report \cite{Video-Streaming-Report} predicts that video will account for 80\% of mobile traffic by 2029. The rapid growth of User-Generated Content (UGC) and Video-on-Demand (VoD) services from platforms such as Netflix, Meta, YouTube, and Prime Video has necessitated the development of adaptive delivery systems that address diverse network conditions and user preferences. However, delivering high-quality videos at scale imposes immense computational demands, particularly during the construction of rate-quality (RQ) curves, a process that requires multiple iterations of compression and quality estimation. While the majority of existing works \cite{Fast-Encoding-Parameter-Selection-For-Convex-Hull-Video-Encoding,Efficient-Bitrate-Ladder-Construction-for-Content-Optimized-Adaptive-Video-Streaming,Benchmarking-Learning-based-Bitrate-Ladder-Prediction-Methods-for-Adaptive-Video-Streaming,Perceptually-Aware-Per-Title-Encoding-for-Adaptive-Video-Streaming,Constructing-Per-Shot-Bitrate-Ladders-using-Visual-Information-Fidelity,Leveraging-Compression-to-Construct-Transferable-Bitrate-Ladders} focus on predicting optimal encoding parameters for individual videos, recent industry efforts by Meta \cite{Towards-Perceptually-Optimized-Compression-Of-User-Generated-Content} and YouTube \cite{Rate-Distortion-Optimization-Over-Large-Scale-Video-Corpus-With-Machine-Learning} have proposed clustering videos by their RQ characteristics, and then applying uniform encoding parameters within these clusters to significantly reduce computational cost. However, their approaches rely on hand-crafted features and lack comprehensive evaluation across datasets. We refer to this problem of clustering videos based on their encoding complexity as \textbf{Video Encoding Complexity Clustering}. This problem is challenging as it requires understanding the intricate relationship between video content, encoding parameters, and the resulting rate-quality characteristics. Hence, there is a need for a more robust and generalizable solution that can effectively capture the encoding complexity of videos across different domains. To the best of our knowledge, there has been no prior work that has systematically evaluated the performance of existing state-of-the-art (SoTA) image/video encoders for this problem. We begin our work by asking the following questions:

\noindent\textbf{Does semantic similarity capture encoding complexity?} Figure \ref{fig:semantics_vs_encoding_complexity} visualizes the RQ curves of videos in the YouTube-UGC dataset \cite{YouTube-UGC} clustered using their category labels and cluster labels obtained from Meta's clustering algorithm \cite{Towards-Perceptually-Optimized-Compression-Of-User-Generated-Content}. It may be observed that clustering videos by their semantic category labels results in highly heterogeneous RQ curves grouped together, while Meta's clustering algorithm yields tightly homogeneous clusters. Specifically, two videos having similar semantic content (e.g., two videos of the `HDR' or `Vlog' category) may or may not have similar RQ curves and can share similarities with a completely different category of videos (e.g., `Animation' or `Gaming'). Hence, semantic similarity is a poor proxy for encoding complexity and motivates the need for a method that can capture the underlying encoding characteristics of videos.

\noindent\textbf{Why can't existing quality assessment models predict encoding complexity?}
Image and Video Quality Assessment (IQA/VQA) \cite{SSIM, VIF, DLM, VMAF, ST-RRED} models are grounded in principles of human visual perception and are designed to measure deviations in a visual signal for evaluating perceptual quality. They have demonstrated remarkable success in various applications, including video compression, streaming, and enhancement. However, these models are primarily designed to assess the distortions in a video, but not to capture the encoding complexity of a video. For example, two videos with similar perceptual quality scores may have vastly different rate-quality profiles, indicating that they require different encoding parameters to achieve the same level of perceptual quality. This disconnect between perceptual assessment and encoding requirements highlights the possible limitations of existing IQA/VQA models in addressing this problem.

\noindent\textbf{Can self-supervised learning capture encoding complexity?}
Self-supervised learning (SSL) has emerged as a powerful paradigm for extracting rich representations from unlabeled data \cite{SimCLR, BYOL, MoCo, DINOv2, Video-MAE, V-JEPA}. These methods have shown remarkable success in various computer vision tasks, including image classification, object detection, and video understanding. However, the application of these models to the problem of video encoding complexity clustering is an open question. While SSL methods can learn rich representations from videos, they may not necessarily capture the specific features that are relevant for encoding complexity. Furthermore, many existing SSL pipelines are explicitly trained to be invariant to spatial and temporal transformations (e.g., cropping, blurring, or jittering), which may not be suitable for encoding complexity prediction, as these transformations can alter the encoding complexity of a video. Also, unlike problems like video classification, video retrieval, and video summarization, which prioritize semantic understanding of the content, encoding complexity prediction requires understanding the intricate relationship between textures, motion, and other low-level spatio-temporal features of the video. Consequently, it remains unclear whether off-the-shelf SSL methods can effectively preserve these low-level features and learn representations suitable for encoding complexity prediction.

We show that existing state-of-the-art IQA, VQA, and image-based SSL methods are fundamentally misaligned for video encoding complexity clustering. While recent video SSL approaches demonstrate improvements, the overall performance remains sub-optimal for real-world applications. To bridge this gap, we introduce \textbf{Compression Echo Contrastive Learning (CECL)}, a novel self-supervised pretraining framework explicitly designed for video encoding complexity. CECL leverages the `Compression Echo' --- the response of a video to compression --- as a supervisory signal to learn robust representations during pretraining. We formulate the contrastive learning objective to directly capture relative spatio-temporal degradation features rather than memorizing arbitrary absolute error thresholds, allowing the model to learn a nuanced understanding of encoding complexity. We show that with only $\approx$1\% additional compute over V-JEPA \cite{V-JEPA}, our method demonstrates improved performance over existing visual encoders across multiple datasets and delivers significant bitrate and quality savings.

%% file: sections/2_related_works.tex
\section{Related Works}
\label{sec:related_works}
In this section, we provide a brief overview of the existing literature on image/video quality assessment, self-supervised learning for image/video understanding, and video encoding complexity clustering. 

Image and Video Quality Assessment (IQA/VQA) models \cite{SSIM, VIF, DLM, VMAF, ST-RRED, CONTRIQUE, CONVIQT, DOVER} have been widely used to evaluate the perceptual quality of images and videos. Recent state-of-the-art models rely on complex deep neural networks to learn rich representations of visual content. CONTRIQUE \cite{CONTRIQUE} employs contrastive learning over images with synthetic and authentic distortions to achieve state-of-the-art performance on various IQA benchmarks. DOVER \cite{DOVER} leverages a transformer-based architecture with a multi-branch design to capture both the aesthetic and technical aspects of videos. These models are designed for perceptual quality assessment and do not capture the encoding complexity characteristics of videos.

Self-supervised learning (SSL) has emerged as a powerful paradigm for learning representations from unlabeled data, and has been successfully applied to various computer vision tasks. CLIP \cite{CLIP} employs a contrastive learning objective to learn joint representations of images and text. DINOv2 \cite{DINOv2} leverages a self-distillation approach to learn robust visual representations without requiring any labeled data. VideoMAE \cite{Video-MAE} employs a masked autoencoder architecture to learn video representations by reconstructing masked spatiotemporal patches, while V-JEPA \cite{V-JEPA} utilizes a joint-embedding architecture to learn representations by predicting abstract latent features of unseen video segments. These methods primarily learn semantic representations, often using augmentations that force invariance to spatial and temporal transformations \cite{SimCLR, BYOL, MoCo, MoCov3}. This invariance may not be suitable for encoding complexity prediction, as these transformations inherently alter the actual encoding complexity of a video.

The problem of video encoding complexity clustering was initially introduced and explored by researchers at Meta \cite{Towards-Perceptually-Optimized-Compression-Of-User-Generated-Content} and YouTube \cite{Rate-Distortion-Optimization-Over-Large-Scale-Video-Corpus-With-Machine-Learning}. The researchers at Meta modeled this complexity problem \cite{Towards-Perceptually-Optimized-Compression-Of-User-Generated-Content} by clustering videos based on their RQ characteristics (Section \ref{sec:preliminaries}) and training a classifier on low-level, handcrafted features to predict the cluster labels of new videos. Building on this foundation, researchers from YouTube \cite{Rate-Distortion-Optimization-Over-Large-Scale-Video-Corpus-With-Machine-Learning} proposed a bitrate allocation method to minimize the average bitrate subject to constraints on average and minimum quality. To achieve this, they trained a cluster-prediction classifier using features extracted from the AV1 passlog. Our work shares the motivation of these methods, but focuses on developing a novel self-supervised learning framework for pretraining video encoders for the downstream task of encoding complexity clustering, along with a comprehensive evaluation against existing state-of-the-art visual encoders.

%% file: sections/3_motivation.tex
\section{Preliminaries}
\label{sec:preliminaries}

The authors of \cite{Towards-Perceptually-Optimized-Compression-Of-User-Generated-Content} proposed a variant of the \textit{K}-Means algorithm that clusters videos based on their RQ curves, utilizing BD-Rate and BD-VMAF values \cite{BD-Metric} as the distance metric. BD-Rate measures the percentage change in bitrate required to achieve the same quality between two videos, while BD-VMAF measures the change in perceptual quality (VMAF score) at the same bitrate. The algorithm iteratively assigns videos to clusters based on their RQ characteristics and updates cluster centroids until convergence.

In our work, we slightly modify the BD-Rate estimation used in the algorithm to make the distance metric symmetric. Specifically, we map the percentage bitrate change back to an exponential scale to ensure the distance metric remains strictly symmetric between any given pair of videos. The details of this modified BD-Rate calculation are provided in the Supplementary Material. We employ this symmetric adaptation throughout our experiments to evaluate downstream performance, and for simplicity, we refer to it by its original name, Meta's clustering algorithm. Following the original work \cite{Towards-Perceptually-Optimized-Compression-Of-User-Generated-Content}, we set the number of clusters to 14 for all our experiments.

%% file: sections/4_methodology.tex
\begin{figure*}
    \centering
    \begin{subfigure}[b]{0.65\linewidth}
        \centering
        \includegraphics[width=\linewidth]{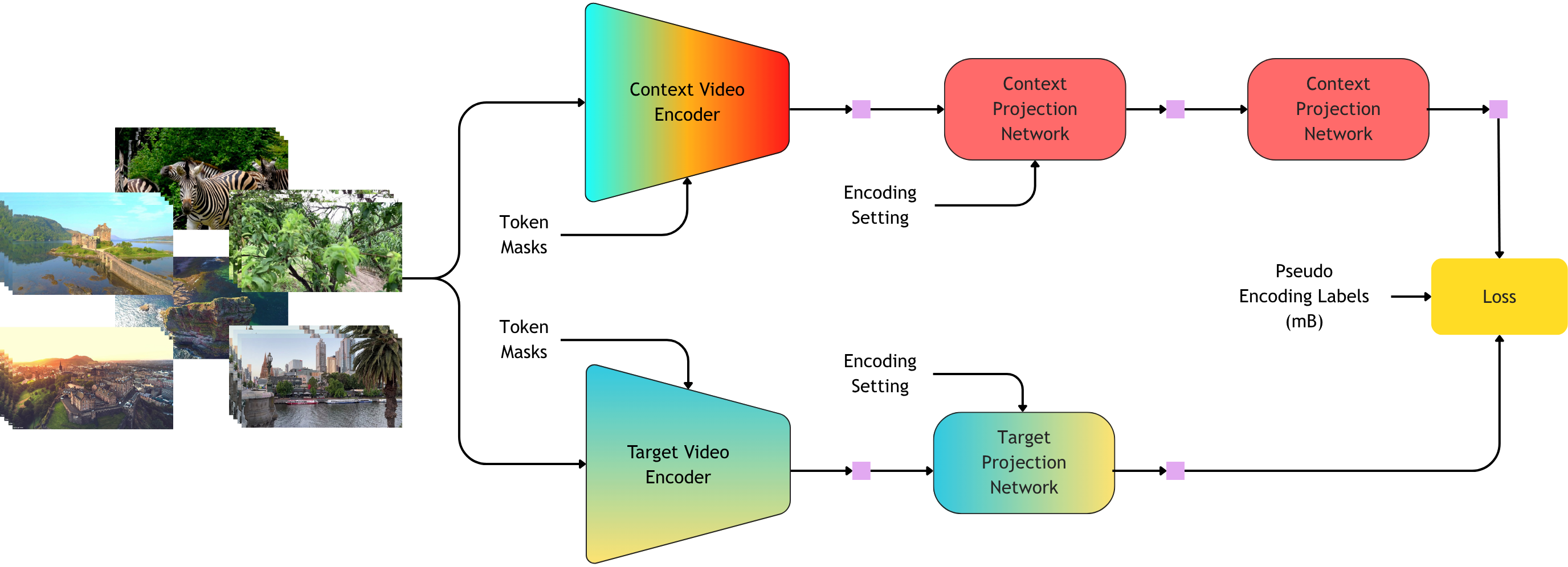}
        \caption{}
        \label{fig:pretraining_method_1}
    \end{subfigure}\hfill
    \begin{minipage}[b]{0.33\linewidth}
        \centering
        \begin{subfigure}[b]{\linewidth}
            \centering
            \includegraphics[width=0.8\linewidth]{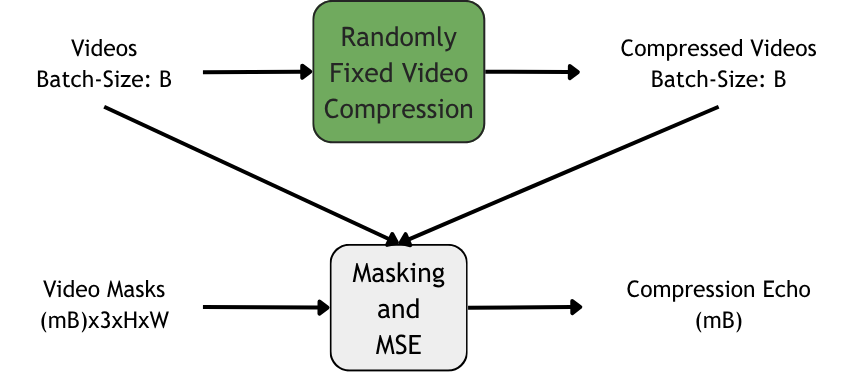}
            \caption{}
            \label{fig:pretraining_method_2}
        \end{subfigure}\\[1em]
        \begin{subfigure}[b]{\linewidth}
            \centering
            \includegraphics[width=0.9\linewidth]{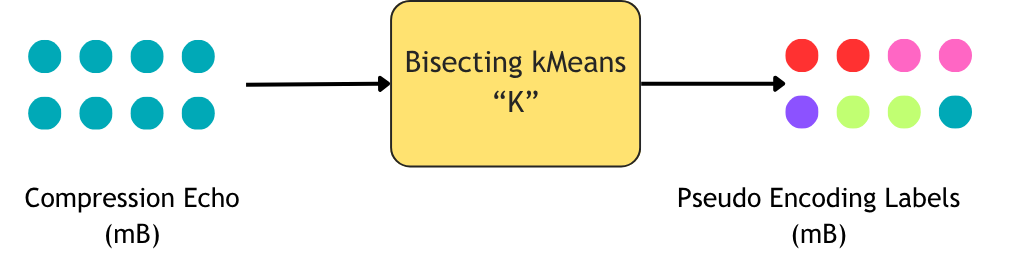}
            \caption{}
            \label{fig:pretraining_method_3}
        \end{subfigure}
    \end{minipage}
    \caption{Overview of the proposed pretraining method for learning video representations for encoding complexity clustering. (a) The pretraining method consists of context and target branches. The target branch is updated using an exponential moving average of the context branch. (b) The compression echo is computed by calculating the MSE between the masked original video and the corresponding masked compressed video. (c) Pseudo-labels are generated by clustering the compression echo values using Bisecting K-Means clustering. Videos within the same cluster are treated as positive pairs, while videos in different clusters are treated as negative pairs.}
    \label{fig:Pretraining}
\end{figure*}

\section{Method}
\label{sec:method}

Based on the observations outlined in the Introduction, we hypothesize that the target features should capture the specific spatio-temporal characteristics most susceptible to video compression. Guided by this insight, we propose a novel self-supervised learning framework for pretraining the video encoder, which we refer to as \textbf{Compression Echo Contrastive Learning (CECL)}. The key idea behind CECL is to leverage the responses of videos to compression as a supervisory signal to learn strong spatio-temporal representations. Our central premise is that videos exhibiting similar compression responses inherently share similar spatio-temporal characteristics. This is because compression algorithms fundamentally exploit spatial and temporal redundancies: videos amenable to high compression efficiency typically contain predictable patterns, while those with low efficiency contain complex motions and spatial details. We formally define the pretraining task as follows:

\paragraph{Batch Formulation and Masking.}
Given an input batch of $B$ unique source videos, we first compress all the videos by choosing a random encoding setting (resolution and CRF value) to obtain their corresponding compressed videos. Let $\text{V}_{i}$ and $\text{V}_{i}^{e_{k}}$ denote the original and compressed versions of the $i^{\text{th}}$ video under the $k^{\text{th}}$ encoding setting, respectively. Since decoding each video during dataloading is a known I/O bottleneck, following \cite{OmniMAE}, we repeat each video $m$ times in the batch and apply a random spatio-temporal mask $M_{i}^{j}$ with a constant masking ratio $r$ each time independently to obtain $m$ masked versions of the original and compressed videos, denoted by $\text{V}_{i}^{j} = \text{M}_{i}^{j} \odot \text{V}_{i}$ and $\text{V}_{i}^{e_{k}, j} = \text{M}_{i}^{j} \odot \text{V}_{i}^{e_{k}}$, respectively, where $j \in \{1, 2, \ldots, m\}$.

\paragraph{The Compression Echo.}
We measure the effect of compression on a video by calculating the Mean Squared Error (MSE) between the masked original video and the corresponding masked compressed video. We formally define this difference as the \textbf{Compression Echo} ($\text{CE}$). Figure \ref{fig:pretraining_method_2} illustrates the pipeline for computing compression echo. The compression echo effectively isolates how susceptible the visible spatio-temporal region is to the compression process with a certain encoding setting, calculated as:
\begin{align}
	\text{CE}^{j}_{i} = \text{MSE}(\text{V}_{i}^{j}, \text{V}_{i}^{e_{k}, j}).
\end{align}

\paragraph{Pseudo Encoding Labels.}
Figure \ref{fig:pretraining_method_3} illustrates an overview of our pseudo encoding label generation process. We hypothesize that videos with similar compression echo values share similar spatio-temporal characteristics, and thus, we can use the compression echo values to generate pseudo-labels for contrastive learning. Specifically, we employ Bisecting K-Means clustering to group the $\text{CE}^{j}_{i}$ values into $K$ clusters, where each cluster represents a pseudo-label. Videos within the same cluster are treated as positive pairs, while videos in different clusters are treated as negative pairs. This clustering process is performed on-the-fly during training, allowing for dynamic generation of pseudo-labels based on the sampled batch of videos, the applied masks, and the sampled encoding setting. 

\paragraph{Optimization Objective.}
Once the pseudo-labels are generated, the masked compressed videos ($\text{V}_{i}^{e_{k}, j}$) are discarded, as they are used solely to compute the supervisory signal. The augmented batch of $mB$ masked source videos ($\text{V}_{i}^{j}$) is then stratified by pseudo-label and split into two halves for context and target branches, respectively. Following \cite{BYOL, MoCov3}, we employ an asymmetrical pretraining architecture with video encoders and projection networks for both branches, while the context branch additionally includes a predictor. Figure \ref{fig:pretraining_method_1} shows the architecture of our proposed pretraining method.

During pretraining, the context video encoder processes its split to produce patch-level representations, upon which a single Global Average Pooling (GAP) operation is performed to derive the global representation. The global representation, along with encoding settings (resolution and CRF), is then passed through a projection network and followed by a prediction network to generate context projections for the contrastive objective. The target projections are generated similarly, but without the prediction network. We employ a contrastive loss function to maximize the similarity between the context and target projections of videos within the same pseudo-label cluster while minimizing the similarity between videos from different clusters:
\begin{align}
	\mathcal{L} = \sum_{i} \frac{-1}{|\mathcal{P}(i)|} \sum_{p \in \mathcal{P}(i)} \log \frac{\exp(\mathbf{q}_i \cdot \mathbf{k}_p / \tau)}{\sum_{j \in \mathcal{P}(i) \cup \mathcal{N}(i)} \exp(\mathbf{q}_i \cdot \mathbf{k}_j / \tau)}
\end{align}
where $\mathcal{P}(i)$ and $\mathcal{N}(i)$ denote the sets of positive and negative samples for the $i^{\text{th}}$ masked context video, respectively. $\mathbf{q}_i$ and $\mathbf{k}_j$ denote the context and target projections, respectively, and $\tau$ is a temperature hyperparameter. We update the parameters of the target networks using an exponential moving average of the parameters of the context networks. 

Our proposed pretraining method is fully self-supervised and does not rely on any external labels, making it applicable to any video dataset. The pseudo-labels derived from the compression echo offer a highly adaptable learning signal that depends only on the sampled batch, encoding setting, applied masks, and the hyperparameter $K$. We compute this signal using MSE as opposed to VMAF or bitrate, since relying on those metrics would pull the pseudo-labels toward the very BD-VMAF and BD-Rate quantities that define our ground-truth labels, blurring the line between the learning signal and the evaluation target. By contrast, those ground-truth labels from Meta's clustering algorithm are far more restrictive, since grouping two videos requires nearly identical BD-Rate and BD-VMAF values across multiple encoding points. Hence, our proposed pretraining method allows for a more flexible and generalizable learning of representations for video encoding complexity prediction.

%% file: sections/5_experimental_setup_datasets.tex
\section{Experimental Setup and Datasets}
\label{sec:experimental_setup_and_datasets}
\subsection{Training and Evaluation Frameworks}
\begin{figure}
	\centering
	\begin{subfigure}{\columnwidth}
		\centering
		\includegraphics[width=\columnwidth]{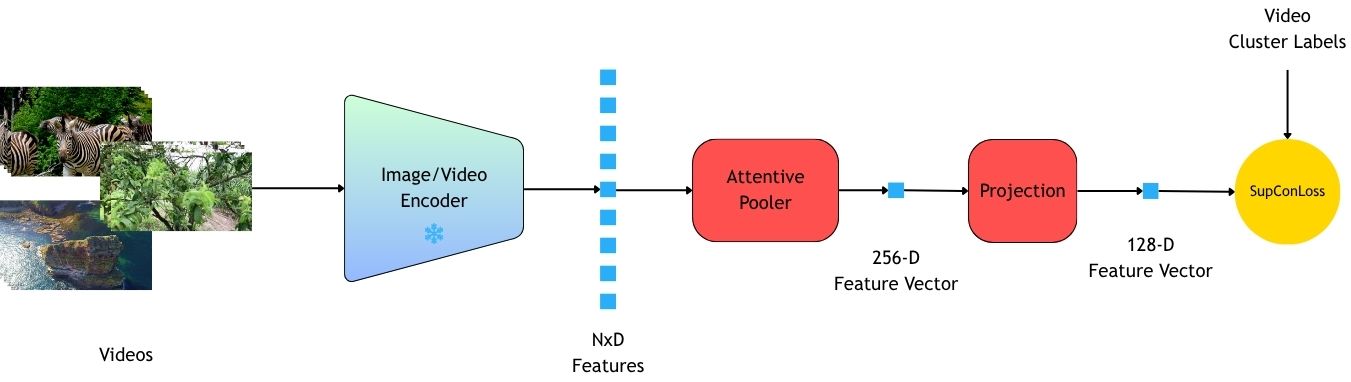}
		\caption{Training Framework}
		\label{fig:training_framework}
	\end{subfigure}\\[1em]
	\begin{subfigure}{\columnwidth}
		\includegraphics[width=\columnwidth]{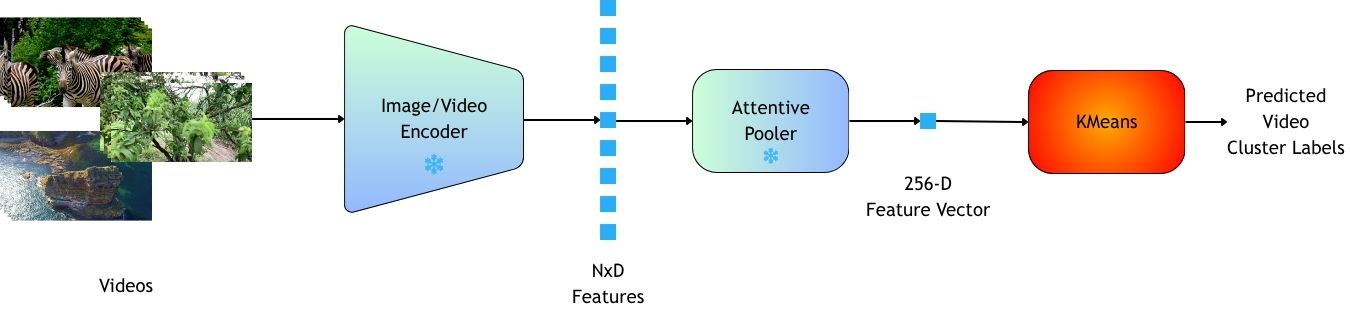}
		\caption{Evaluation Framework}
		\label{fig:evaluation_framework}
	\end{subfigure}
	\caption{Training and Evaluation Frameworks for Video Encoding Complexity Clustering.}
	\label{fig:training_and_evaluation_framework}
\end{figure}

We evaluate our proposed method by measuring its clustering performance on several diverse publicly available datasets. We compared our method against state-of-the-art models from quality assessment: CONTRIQUE \cite{CONTRIQUE}, DOVER \cite{DOVER}, image SSL: CLIP \cite{CLIP}, DINOv2 \cite{DINOv2}, and video SSL: Video-MAE \cite{Video-MAE}, V-JEPA \cite{V-JEPA}. Figure \ref{fig:training_and_evaluation_framework} shows the training and evaluation framework for the video encoding complexity clustering task. We analyze the performance of these visual encoders by freezing the visual encoder and training a small AttentivePooler on top of the visual encoder to cluster the video representations. 

For our proposed method and the baseline comparisons, we used the ViT-L architecture \cite{Vision-Transformer}. Specifically, we compare against the ViT-L/16 variants for Video-MAE \cite{Video-MAE}, V-JEPA, and the closely related ViT-L/14 variant for DINOv2 \cite{DINOv2} and CLIP \cite{CLIP}. As established in Section \ref{sec:method}, pretraining the video encoder should be carried out in native resolution, as resizing the videos for the pretraining may alter the fundamental properties of the videos that define their response to compression. This, unfortunately, causes a huge data constraint for the pretraining, as we cannot leverage large-scale video datasets that are available at lower resolutions. Since it is not feasible to run compression during pretraining, we precompute compressed videos for pretraining. However, this precomputation limits the data-augmentation strategies that can be applied to the videos during pretraining, as most of the augmentations can change the encoding complexity of videos, and hence, we restrict the data-augmentation strategies to limited random spatial and temporal cropping and random masking.

\subsection{Datasets and Splits}
\label{sec:datasets}
We employed four publicly available datasets in our experiments: LAVIB \cite{LAVIB}, OpenVid \cite{OpenVid}, Inter4K \cite{Inter4K}, and YouTube-UGC \cite{YouTube-UGC}. LAVIB was used when pretraining the video encoder, while both LAVIB and OpenVid served as training and validation sets for attention-probing. Inter4K and YouTube-UGC datasets were utilized exclusively for validation to assess cross-domain generalization. Table \ref{tab:datasets} summarizes the datasets, their purposes, resolutions, and the number of videos in the training and validation splits. Each dataset has its own unique characteristics based on its source, and hence, we applied tailored preprocessing pipelines to extract high-quality video scenes from each video in the dataset. The detailed preprocessing steps for each dataset are described in the Supplementary Material. To ensure balanced cluster representations, we stratified training and validation splits based on cluster labels generated by Meta's Video Complexity Clustering algorithm \cite{Towards-Perceptually-Optimized-Compression-Of-User-Generated-Content}, using an 80:20 ratio. The cluster labels were separately regenerated for each split to avoid bias. To strictly prevent data leakage, the pretraining corpus is restricted exclusively to the LAVIB training split, and all validation sets remain isolated across all training phases.

\subsection{Rate-Quality Points}
We employed \textit{FFmpeg} to perform compression and quality estimation of compressed videos. We deployed Video Multi-Method Assessment Fusion (VMAF) \cite{VMAF} as a quality predictor when evaluating the perceptual quality of the compressed videos. We compressed each video using the libx265 codec (H.265) with the veryfast preset. When pretraining, LAVIB videos were compressed at 720p, 540p, and 432p, while applying seven CRF values (18, 22, 26, 30, 34, 38, 42). For training and evaluation, we compressed videos following the resolution settings detailed in Table \ref{tab:datasets} across the same seven CRF values. The compressed videos employed during pretraining require 2.3 TB of storage, while the source videos across all experiments require 0.75 TB. We observed a similar median execution time for RQ points for each video as described in the work \cite{Leveraging-Compression-to-Construct-Transferable-Bitrate-Ladders}.

\begin{table}[!t]
	\centering
	\caption{Datasets used for Pretraining, Training, and Evaluation.}
	\footnotesize
	\setlength{\tabcolsep}{3pt}
	\resizebox{\columnwidth}{!}{
	\begin{tabular}{lccc}
		\toprule
		Dataset & Purpose & Resolution & \#Videos (Train / Val) \\
		\midrule
		LAVIB & Pretraining, Training/Validation & 720p & 32,000 / 8,000 \\
		OpenVid & Training/Validation & 1080p & 32,000 / 8,000 \\
		Inter4K & Validation (OOD) & 720p, 1080p & - / 1393 \\
		YouTube-UGC & Validation (OOD) & 720p, 1080p & -/ 650 \\
		\bottomrule
	\end{tabular}
}
\label{tab:datasets}
\end{table}

\subsection{Training and Evaluation Settings}
\label{sec:training_and_evaluation_settings}
\textbf{Pretraining:} Since we pretrain the video encoder on a limited amount of data, we initialized it with the weights of a ViT-L/16 with V-JEPA \cite{V-JEPA} model pretrained on the VideoMix2M dataset. We employed random spatial cropping to a resolution of 224$\times$224 and random temporal cropping to 16 frames during pretraining. For pretraining, we set the number of masks $m$ to 4, the masking ratio $r$ to 0.9 \cite{Video-MAE, V-JEPA} with random masking, $K$ to 10, the output of projection and predictor networks \cite{MoCov3} is set to 128, and the local source batch size $B$ to 96. The loss was computed on $mB$ masked videos in each batch, whose pseudo labels are calculated after compression under the sampled encoding setting for that step. We trained our model over 100 epochs using distributed training with an effective global batch size of 768 source videos across all GPUs. While we initialized the video encoder with V-JEPA \cite{V-JEPA} weights, it should be noted that our overall pretraining cost is about only 1.19\% of additional pretraining on top of the V-JEPA \cite{V-JEPA} pretraining. Hence, any gains in performance from our pretraining method are not due to the additional pretraining cost, but rather due to the effectiveness of our proposed pretraining method. The detailed optimization hyperparameters, including momentum scheduler, learning rate, and weight decay schedules, are provided in the Supplementary Material. 

\noindent\textbf{Training:} We trained the networks using a Supervised Contrastive Loss \cite{Supervised-Contrastive-Learning}. We resize the videos to 224$\times$224 spatial resolution and randomly sample 16 frames from the videos during training. We trained our models over 10 epochs using a batch size of 96. Additional details about the training settings, including optimization hyperparameters and the architecture of the AttentivePooler, are provided in the Supplementary Material.

\noindent\textbf{Evaluation:} We evaluated the trained models by extracting the representations for each video, followed by L2 normalization, then applying \textit{K}-Means clustering to cluster the videos together. We measured the clustering performance between the predicted cluster labels and the ground truth cluster labels using standard performance metrics, including Normalized Mutual Information (NMI), Adjusted Rand Index (ARI), and Fowlkes-Mallows Index (FMI). The ground truth cluster labels are obtained using Meta's Video Complexity Clustering algorithm \cite{Towards-Perceptually-Optimized-Compression-Of-User-Generated-Content}. We report the average clustering performance across 10 different random seeds of \textit{K}-Means clustering (set to 14, similar to \cite{Towards-Perceptually-Optimized-Compression-Of-User-Generated-Content}) to ensure the robustness of our results. We evaluate the real-world applicability of our method by measuring BD metrics \cite{BD-Metric} performance against the Fixed Bitrate Ladder \cite{Fixed-Bitrate-Ladder}.

%% file: sections/6_experiments_results.tex
\begin{table}
	\centering
	\footnotesize
	\setlength{\tabcolsep}{2.5pt}
	\caption{Clustering performance of various frozen visual encoders trained on the LAVIB dataset, evaluated against ground-truth complexity labels extracted from 1280$\times$720 videos.}
	\label{tab:frozen_encoder_results_720p}
	\renewcommand{\arraystretch}{1.15}
	\resizebox{\columnwidth}{!}{
	\begin{tabular}{lccccccccc}
		\toprule
		& \multicolumn{3}{c}{In-Domain} & \multicolumn{6}{c}{Out-of-Domain} \\
		\cmidrule(lr){2-4} \cmidrule(lr){5-10}
		Method & \multicolumn{3}{c}{LAVIB} & \multicolumn{3}{c}{Inter4K} & \multicolumn{3}{c}{YouTube-UGC} \\
		\cmidrule(lr){2-4} \cmidrule(lr){5-7} \cmidrule(lr){8-10}
		& ARI & NMI & FMI & ARI & NMI & FMI & ARI & NMI & FMI \\
		\midrule
		CONTRIQUE \cite{CONTRIQUE} & 0.036 & 0.097 & 0.110 & 0.027 & 0.093 & 0.104 & 0.045 & 0.166 & 0.122 \\
		DOVER \cite{DOVER} & 0.106 & 0.276 & 0.178 & 0.053 & 0.189 & 0.133 & 0.053 & 0.198 & 0.130 \\
		\midrule
		CLIP \cite{CLIP} & 0.080 & 0.189 & 0.154 & 0.034 & 0.124 & 0.111 & 0.032 & 0.147 & 0.111 \\
		DINOv2 \cite{DINOv2} & 0.060 & 0.168 & 0.138 & 0.030 & 0.114 & 0.109 & 0.043 & 0.165 & 0.121 \\
		\midrule
		Video-MAE \cite{Video-MAE} & \underline{0.193} & \underline{0.399} & \underline{0.257} & 0.116 & \underline{0.302} & 0.186 & 0.114 & 0.298 & 0.187 \\
		V-JEPA \cite{V-JEPA} & 0.179 & 0.383 & 0.244 & \underline{0.121} & \underline{0.302} & \underline{0.192} & \underline{0.134} & \underline{0.330} & \underline{0.204} \\
		\rowcolor{cyan!20} Our Method & \textbf{0.226} & \textbf{0.442} & \textbf{0.287} & \textbf{0.128} & \textbf{0.333} & \textbf{0.198} & \textbf{0.147} & \textbf{0.335} & \textbf{0.215} \\
		\bottomrule
	\end{tabular}
}
\end{table}

\begin{table}
	\centering
	\footnotesize
	\setlength{\tabcolsep}{2.5pt}
	\caption{Clustering performance of various frozen visual encoders trained on the OpenVid dataset, evaluated against ground-truth complexity labels extracted from 1920$\times$1080 videos.}
	\label{tab:frozen_encoder_results_1080p}
	\renewcommand{\arraystretch}{1.15}
	\resizebox{\columnwidth}{!}{
   \begin{tabular}{lccccccccc}
		\toprule
		& \multicolumn{3}{c}{In-Domain} & \multicolumn{6}{c}{Out-of-Domain} \\
		\cmidrule(lr){2-4} \cmidrule(lr){5-10}
		Method & \multicolumn{3}{c}{OpenVid} & \multicolumn{3}{c}{Inter4K} & \multicolumn{3}{c}{YouTube-UGC} \\
		\cmidrule(lr){2-4} \cmidrule(lr){5-7} \cmidrule(lr){8-10}
		& ARI & NMI & FMI & ARI & NMI & FMI & ARI & NMI & FMI \\
		\midrule
		CONTRIQUE \cite{CONTRIQUE} & 0.012 & 0.035 & 0.085 & 0.028 & 0.096 & 0.103 & 0.052 & 0.163 & 0.138 \\
		DOVER \cite{DOVER} & 0.014 & 0.040 & 0.091 & 0.011 & 0.056 & 0.092 & 0.021 & 0.103 & 0.115 \\
		\midrule
		CLIP \cite{CLIP} & 0.050 & 0.117 & 0.122 & 0.040 & 0.123 & 0.117 & 0.049 & 0.162 & 0.133 \\
		DINOv2 \cite{DINOv2} & 0.020 & 0.072 & 0.099 & 0.025 & 0.101 & 0.102 & 0.042 & 0.154 & 0.125 \\
		\midrule
		Video-MAE \cite{Video-MAE} & \underline{0.133} & 0.299 & 0.198 & 0.141 & \underline{0.344} & 0.216 & 0.122 & 0.299 & 0.203 \\
		V-JEPA \cite{V-JEPA} & \textbf{0.146} & \underline{0.305} & \textbf{0.211} & \underline{0.143} & 0.328 & \underline{0.217} & \textbf{0.154} & \textbf{0.334} & \textbf{0.244} \\
		\rowcolor{cyan!20} Our Method & 0.131 & \textbf{0.321} & \underline{0.199} & \textbf{0.155} & \textbf{0.363} & \textbf{0.225} & \underline{0.132} & \underline{0.321} & \underline{0.210} \\
		\bottomrule
	\end{tabular}
}
\end{table}

\section{Experiments and Results}
\label{sec:experiments_results}
\begin{figure*}[!ht]
	\centering
	\begin{subfigure}{0.24\textwidth}
		\includegraphics[width=\textwidth]{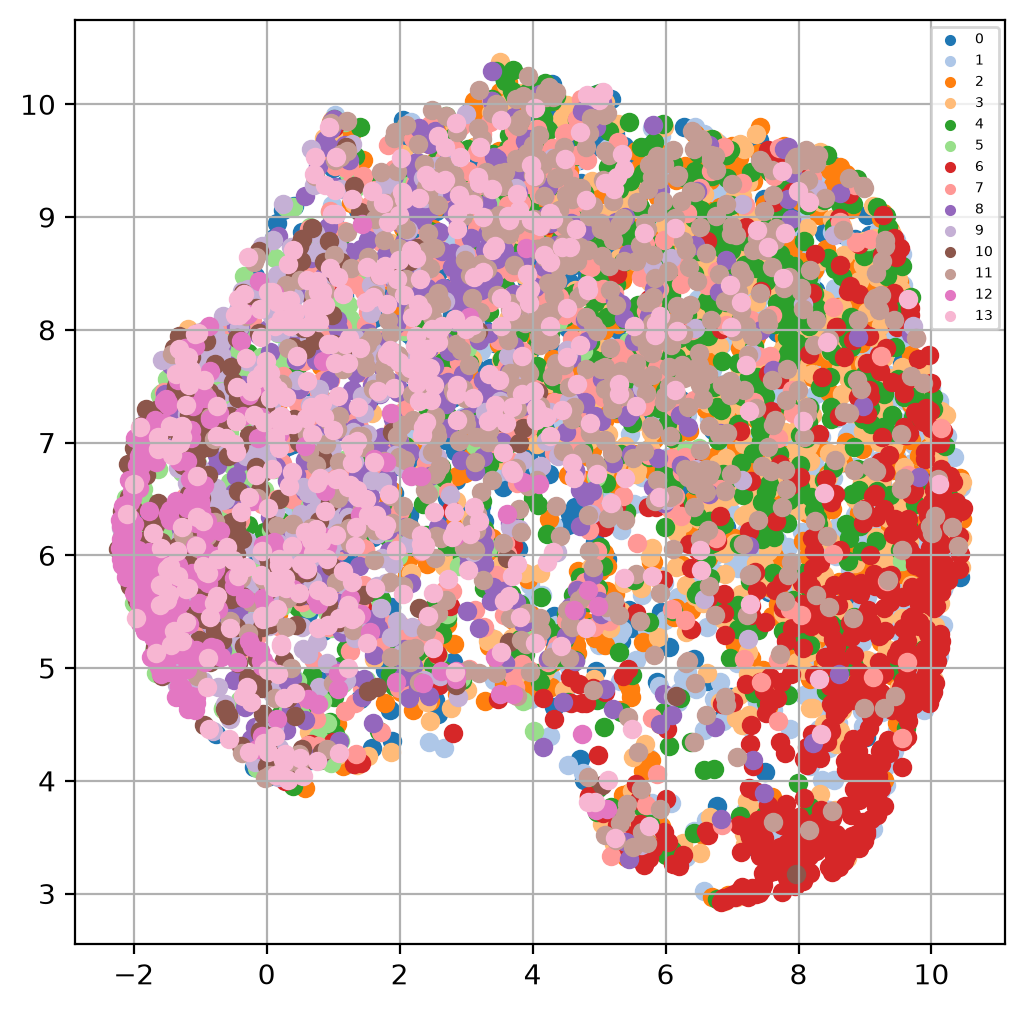}
		\caption{CLIP}
	\end{subfigure}
	\hfill
	\begin{subfigure}{0.24\textwidth}
		\includegraphics[width=\textwidth]{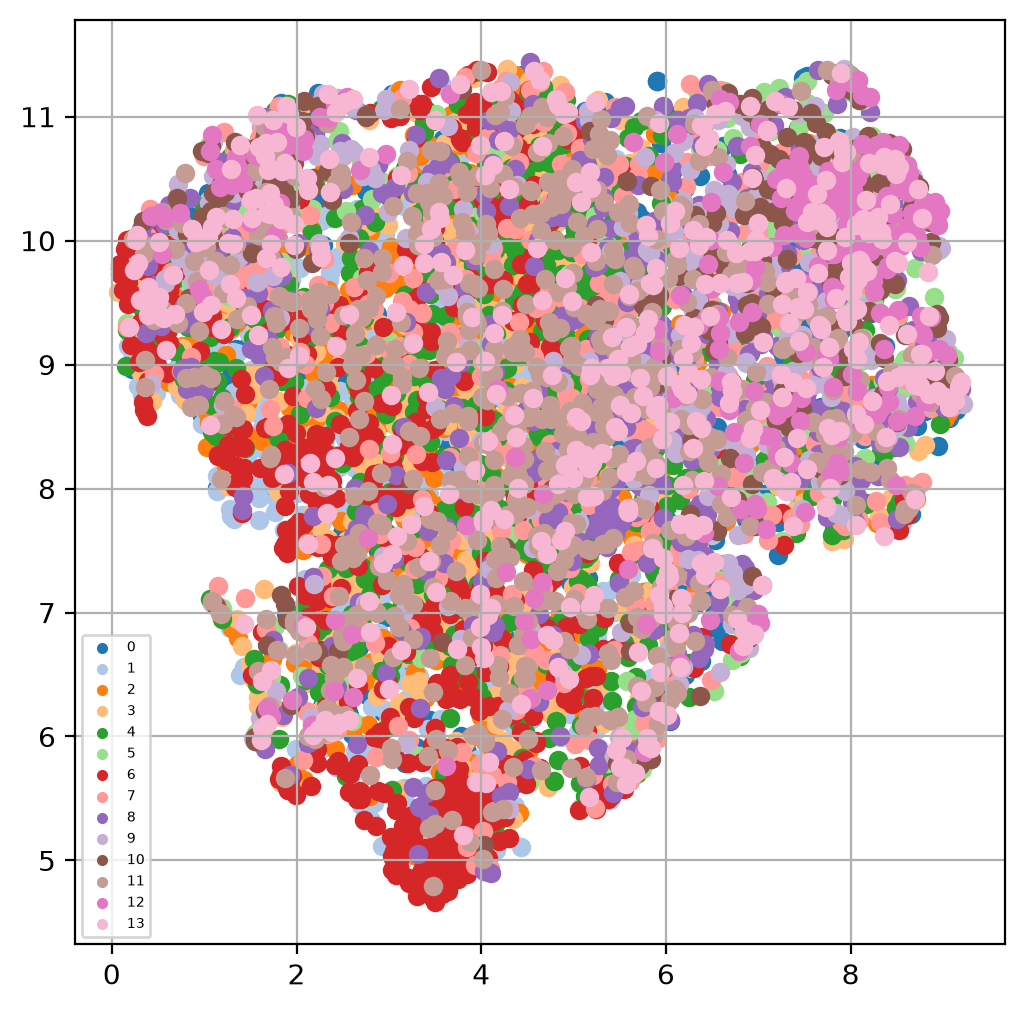}
		\caption{CONTRIQUE}
	\end{subfigure}
	\hfill
	\begin{subfigure}{0.24\textwidth}
		\includegraphics[width=\textwidth]{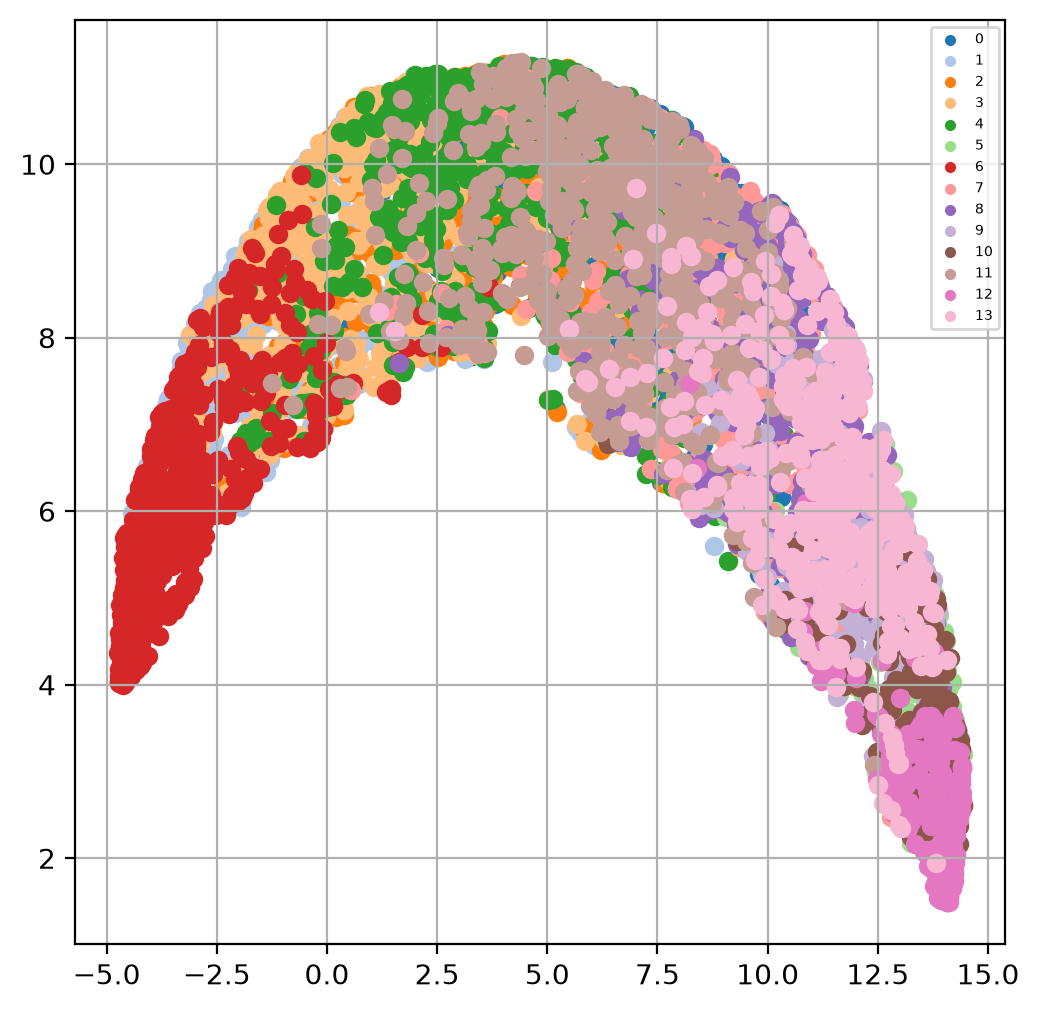}
		\caption{V-JEPA}
	\end{subfigure}
	\hfill
	\begin{subfigure}{0.24\textwidth}
		\includegraphics[width=\textwidth]{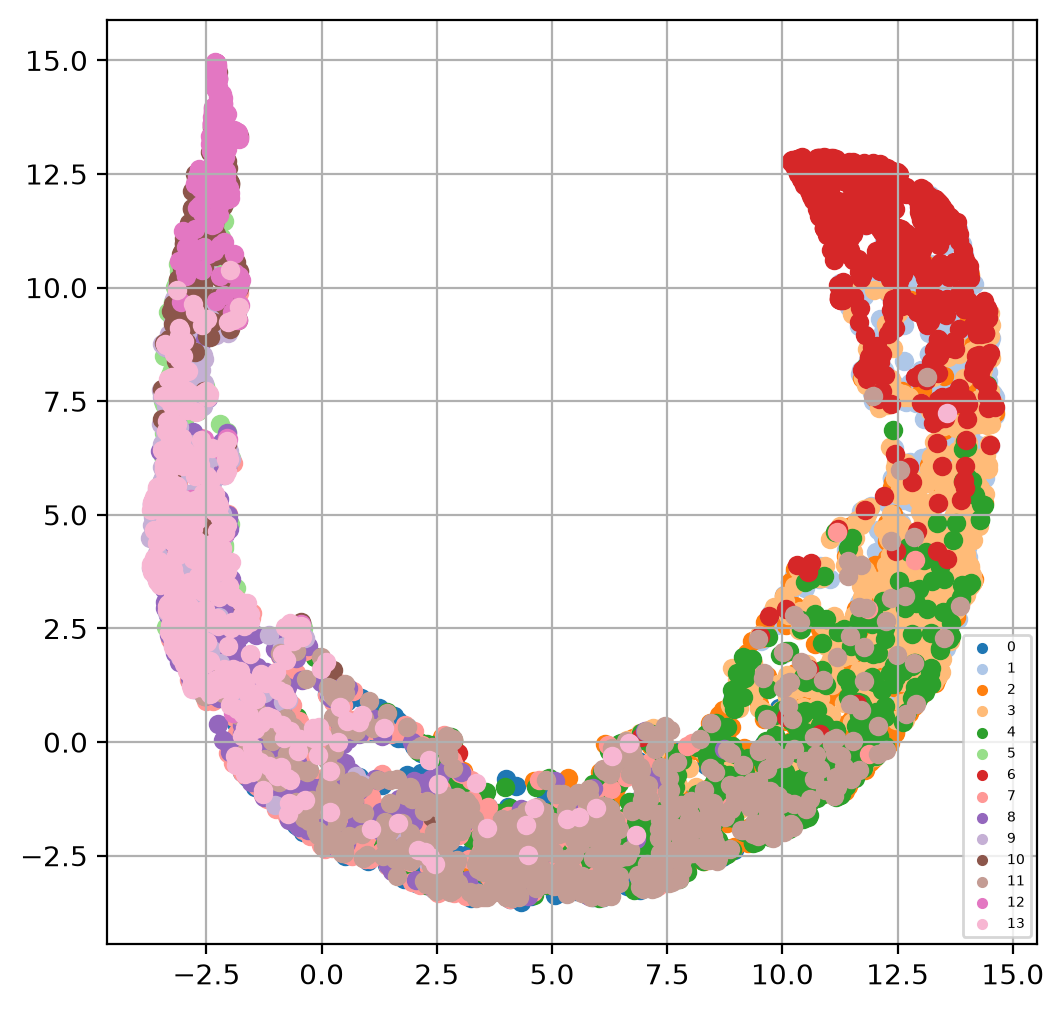}
		\caption{CECL}
	\end{subfigure}
	\caption{UMAP visualizations of the video representations from visual encoders. The models were trained and evaluated on the LAVIB dataset. Each color represents a different cluster label obtained from Meta's Video Complexity Clustering algorithm \cite{Towards-Perceptually-Optimized-Compression-Of-User-Generated-Content}.}
	\label{fig:umap_visualizations}
\end{figure*}

\subsection{Clustering Performance}
\label{sec:clustering_performance}
Tables \ref{tab:frozen_encoder_results_720p} and \ref{tab:frozen_encoder_results_1080p} show the clustering performances of frozen visual encoders trained on LAVIB and OpenVid, then evaluated on ground-truth complexity labels extracted from 1280$\times$720 and 1920$\times$1080 videos, respectively. The performance on 1080p labels evaluates the model's robustness to unseen pretraining resolutions. The tables report the mean clustering performance across three metrics across 10 different random seeds. The best-performing method for each dataset is highlighted in bold, while the second-best is underlined. 

\begin{figure}
	\centering
	\begin{subfigure}{0.475\columnwidth}
		\centering
		\includegraphics[width=\textwidth]{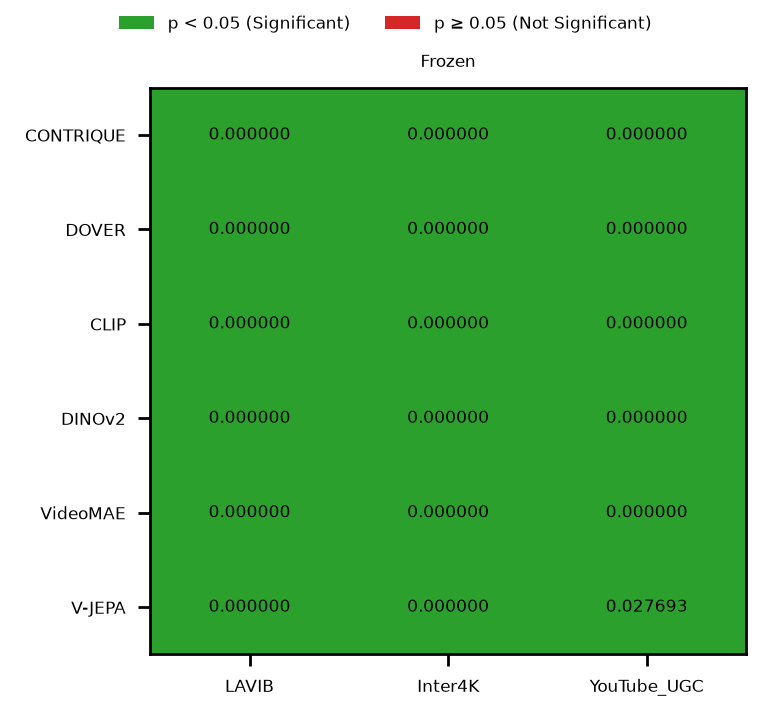}
		\caption{LAVIB training\\1280$\times$720 evaluation}
		\label{fig:t_test_lavib}
	\end{subfigure}
	\hfill
	\begin{subfigure}{0.475\columnwidth}
		\centering
		\includegraphics[width=\textwidth]{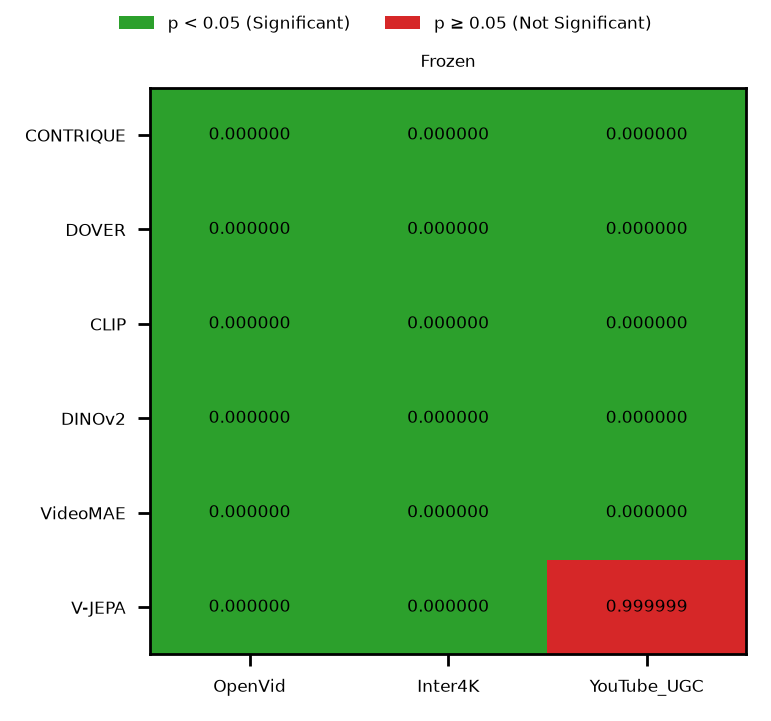}
		\caption{OpenVid training\\1920$\times$1080 evaluation}
		\label{fig:t_test_openvid}
	\end{subfigure}
	\caption{One-sided \textit{t}-test results for our performance gains on NMI against all other methods on LAVIB and OpenVid datasets.}
	\label{fig:t_test_results}
\end{figure}

Among the baselines, it may be observed that encoders like CONTRIQUE \cite{CONTRIQUE}, DOVER \cite{DOVER}, CLIP \cite{CLIP}, and DINOv2 \cite{DINOv2} performed poorly on both in-domain and out-of-domain datasets, while video encoders like Video-MAE \cite{Video-MAE} and V-JEPA \cite{V-JEPA} performed significantly better. On 720p video complexity labels, it may be observed that our method outperformed every baseline on both in-domain and out-of-domain datasets across all three metrics. On 1080p videos, our method achieved the best performance on Inter4K, while V-JEPA \cite{V-JEPA} achieved the best performance on the YouTube-UGC dataset. We validate the statistical significance of our performance gains using a one-sided \textit{t}-test. Figure \ref{fig:t_test_results} shows the $p$-values for our performance gains on NMI against all other methods on LAVIB and OpenVid datasets. It may be observed that our performance gains are statistically significant with $p$-values less than 0.05 against all other methods on both datasets, except for the 1080p YouTube-UGC dataset.

These results indicate that models trained on image invariant features and multimodal features are not sufficient to learn representations of encoding complexity clustering, while video encoders like Video-MAE \cite{Video-MAE} and V-JEPA \cite{V-JEPA} are better suited for this task. Additionally, IQA-based pretraining objectives, as used in CONTRIQUE \cite{CONTRIQUE}, are also ill-suited for this task, as they are designed to learn representations of perceptual quality rather than encoding complexity. These results demonstrate the effectiveness of our proposed pretraining strategy in learning representations of encoding complexity clustering. It should be noted that evaluating the clustering performance of visual encoders only provides a partial view of the overall performance of the models, as it does not take into account the performance in real-world video compression scenarios. Hence, we also evaluate the performance of our method in real-world video compression scenarios using BD metrics \cite{BD-Metric}.

\begin{figure*}
	\centering
	\begin{subfigure}{0.24\textwidth}
		\centering
		\includegraphics[width=\textwidth]{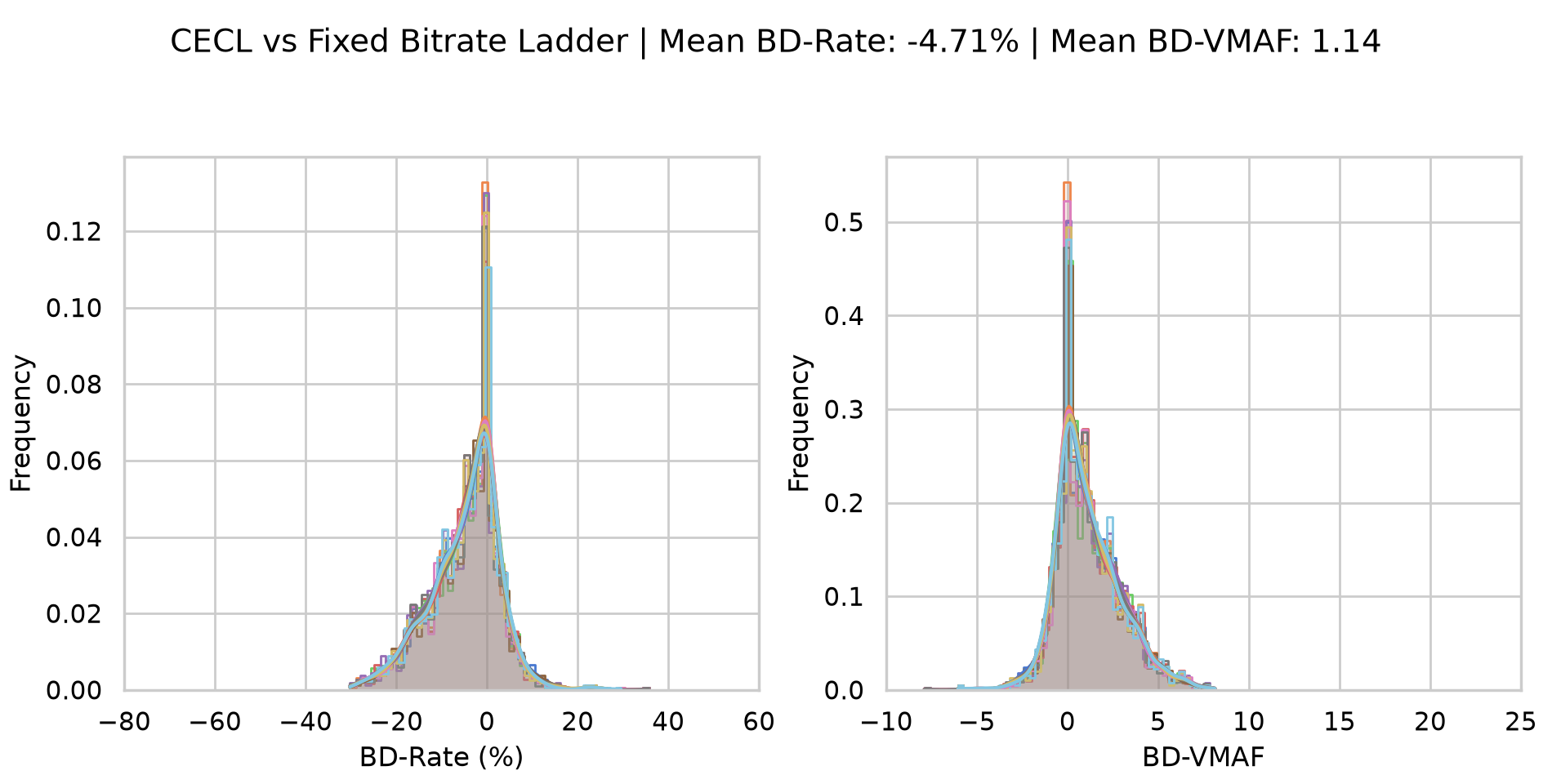}
		\caption{Inter4K, CECL}
	\end{subfigure}
	\hfill
	\begin{subfigure}{0.24\textwidth}
		\centering
		\includegraphics[width=\textwidth]{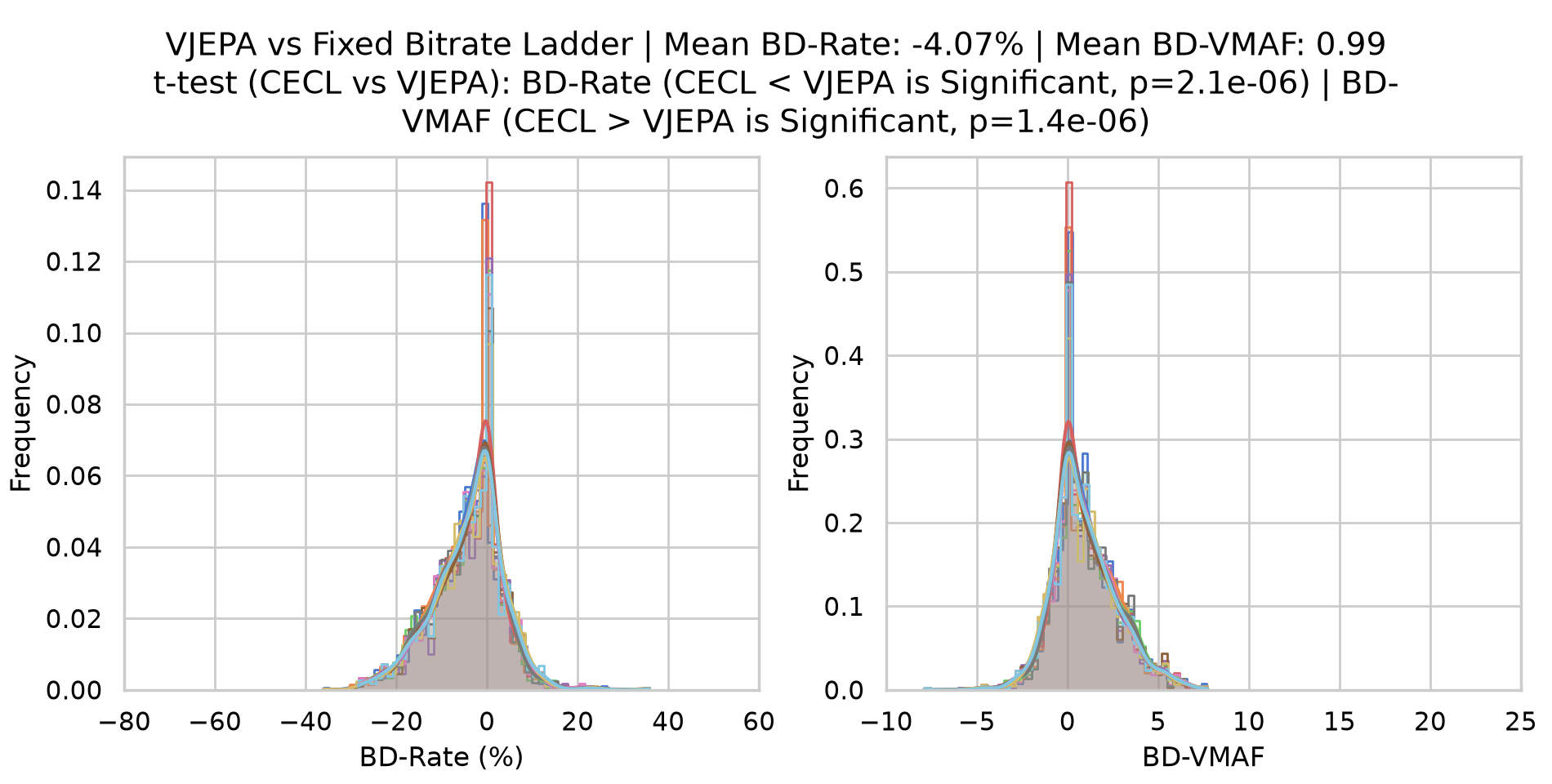}
		\caption{Inter4K, V-JEPA}
	\end{subfigure}
	\hfill
	\begin{subfigure}{0.24\textwidth}
		\centering
		\includegraphics[width=\textwidth]{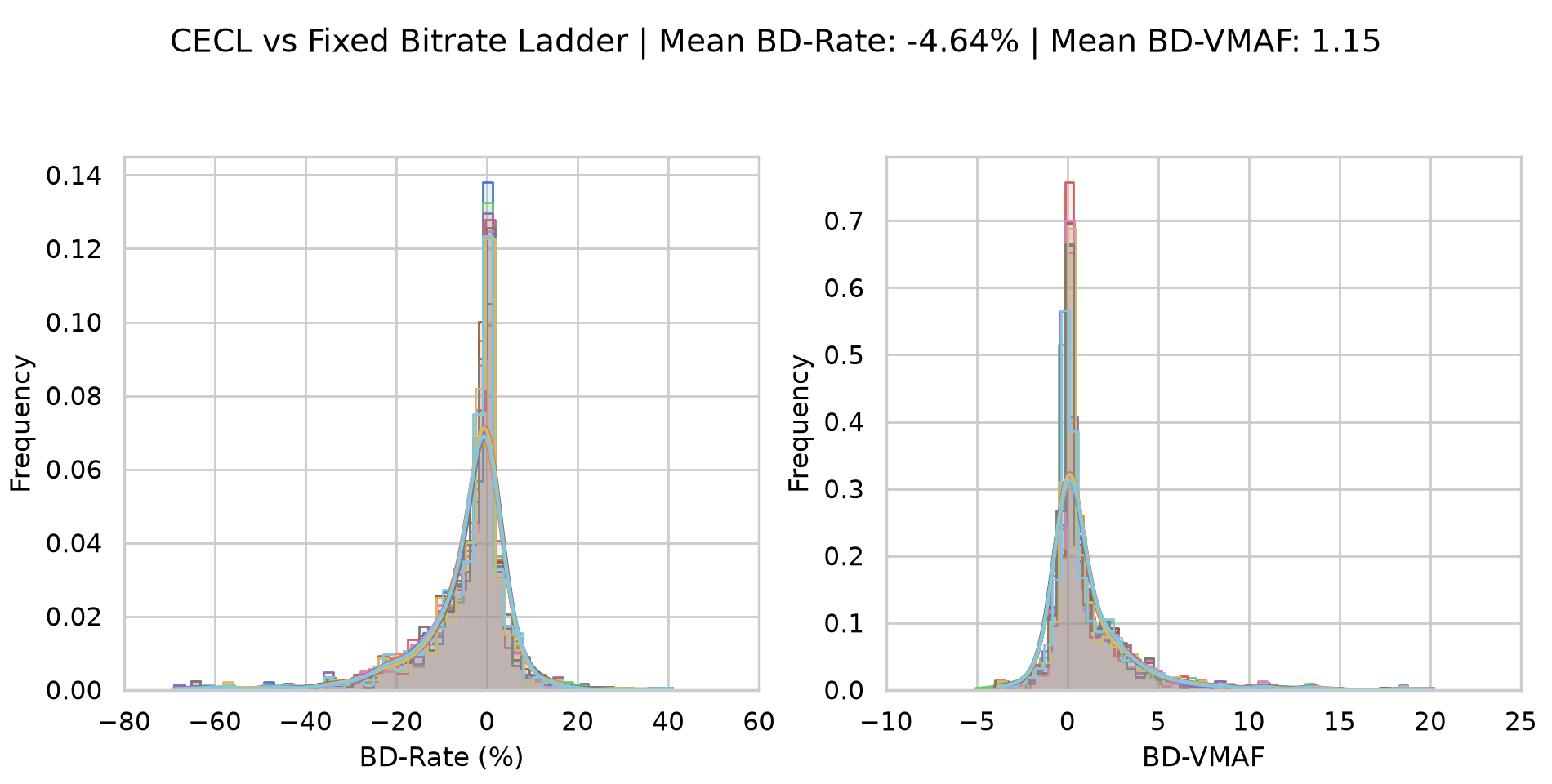}
		\caption{YouTube-UGC, CECL}
	\end{subfigure}
	\hfill
	\begin{subfigure}{0.24\textwidth}
		\centering
		\includegraphics[width=\textwidth]{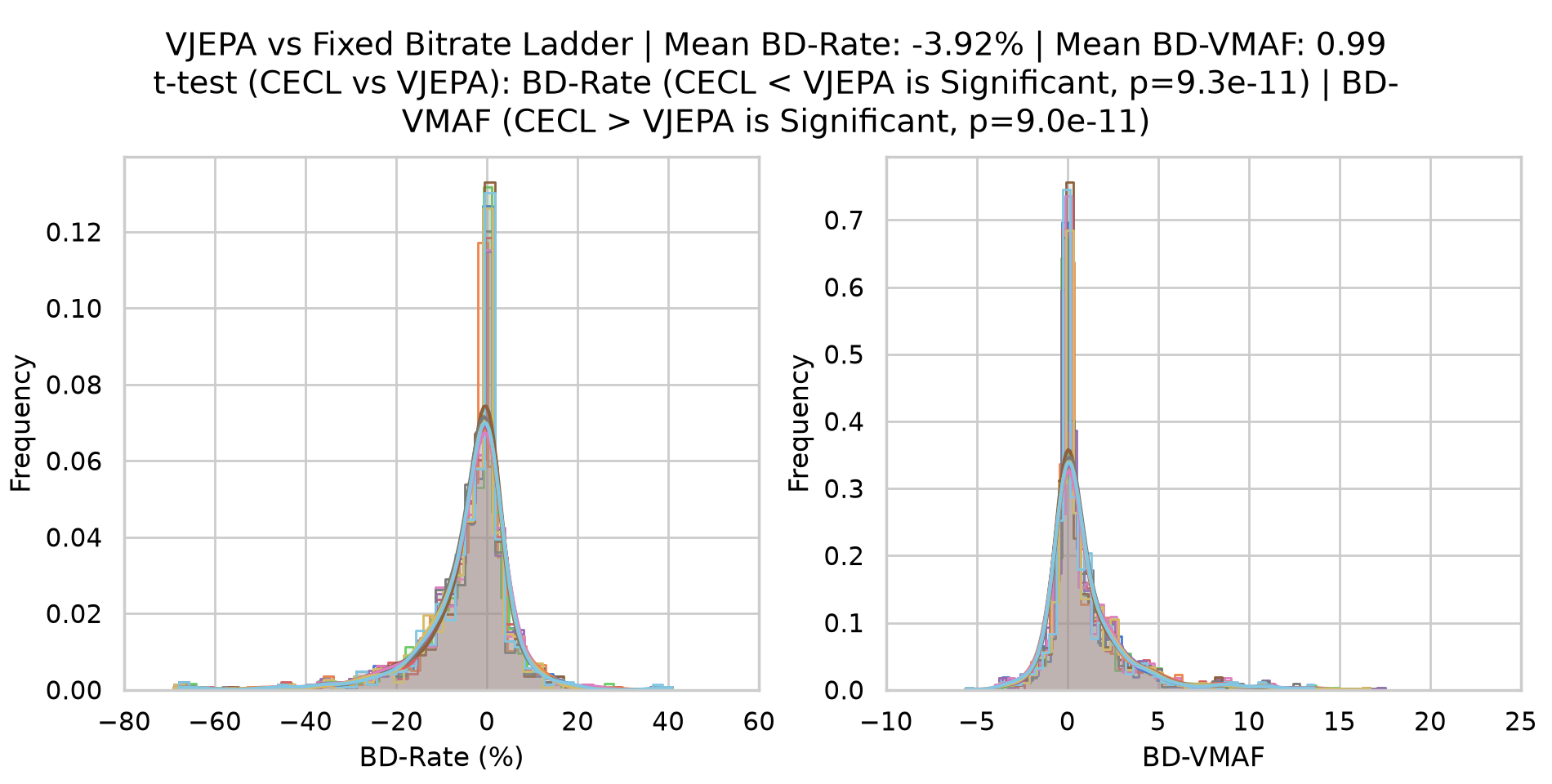}
		\caption{YouTube-UGC, V-JEPA}
	\end{subfigure}
	\caption{BD metrics performance of CECL and V-JEPA on Inter4K and YouTube-UGC datasets using 720p and 1080p videos.}
	\label{fig:encoding_evaluation_histograms}
\end{figure*}

\subsection{Qualitative Analysis}
\label{sec:qualitative_analysis}
Figure \ref{fig:umap_visualizations} shows UMAP visualizations of the video representations from frozen visual encoders trained on LAVIB and evaluated on ground-truth complexity labels extracted from 1280$\times$720 videos. Each color represents a different cluster label obtained from Meta's Video Complexity Clustering algorithm \cite{Towards-Perceptually-Optimized-Compression-Of-User-Generated-Content}. It may be observed that the poor performance of CLIP \cite{CLIP} and CONTRIQUE \cite{CONTRIQUE} is reflected in the poor separation of representations in the UMAP visualizations, while the better performance of V-JEPA \cite{V-JEPA} and our method shows better clustering of the video representations. Comparing the UMAP visualizations of V-JEPA \cite{V-JEPA} and our method, our method shows slightly more concentrated clustering of the video representations, consistent with its stronger quantitative performance on the LAVIB dataset.

\begin{table}
	\centering
	\footnotesize
	\setlength{\tabcolsep}{5pt}
	\caption{Ablation study on the number of clusters ($K$) employed during pretraining.}
	\label{tab:ablation}
	\renewcommand{\arraystretch}{1.15}
	\resizebox{\columnwidth}{!}{
	\begin{tabular}{lccccccc}
		\toprule
		& \multicolumn{3}{c}{Clustering Performance} & \multicolumn{4}{c}{Cluster Sizes} \\
		\cmidrule(lr){2-4} \cmidrule(lr){5-8}
		Method & ARI & NMI & FMI & Min & Max & Median & Mean \\
		\midrule
		CECL ($K = 10$) & \textbf{0.2255} & \textbf{0.4424} & \textbf{0.2868} & 6.0 & 83.6 & 35.3 & 38.4\\
		CECL ($K = 15$) & 0.2203 & 0.439 & 0.2813 & 3.7 & 58.5 & 23.6 & 25.6\\
		CECL ($K = 20$) & 0.2227 & 0.4399 & 0.2834 & 2.5 & 46.2 & 17.5 & 19.2\\
		\bottomrule
	\end{tabular}
}
\end{table}

\subsection{Real-World Evaluation}
\label{sec:real_world_video_compression_evaluation}
We employ the predictions for 720p and 1080p videos from Tables \ref{tab:frozen_encoder_results_720p} and \ref{tab:frozen_encoder_results_1080p}, respectively, to compute per-cluster mean RQ curves at both resolutions. Using these per-cluster mean RQ curves, we derive a predicted cross-over bitrate for each video and use it to construct the video's convex hull. We evaluate the performance of these convex hulls against the Fixed Bitrate Ladder \cite{Fixed-Bitrate-Ladder} using BD metrics \cite{BD-Metric}. The ground-truth labels from Meta's algorithm \cite{Towards-Perceptually-Optimized-Compression-Of-User-Generated-Content} establish an oracle upper bound: a mean BD-Rate of -6.71\% and BD-VMAF of 1.60 for Inter4K and a mean BD-Rate of -5.03\% and BD-VMAF of 1.21 for YouTube-UGC. Figure \ref{fig:encoding_evaluation_histograms} shows the BD-Rate and BD-VMAF of CECL and V-JEPA \cite{V-JEPA} on both datasets, with subfigure titles reporting mean BD values and significance across 10 random seeds. It may be observed that, although our method underperformed V-JEPA in clustering on YouTube-UGC at 1080p, our method outperforms V-JEPA on both datasets in BD-Rate and BD-VMAF metrics, with statistically significant $p < 0.05$ values. On YouTube-UGC, CECL demonstrates BD-Rate of -4.64\% against the -5.03\% oracle and V-JEPA's -3.92\% while on Inter4K, CECL achieves -4.71\% against the -6.71\% oracle and V-JEPA's -4.07\%, recovering a substantial fraction of the achievable headroom. These results indicate that our proposed method CECL is effective in learning encoding-complexity representations that generalize well to real-world video compression scenarios.

\subsection{Ablation Studies}
Table \ref{tab:ablation} shows the in-domain clustering performance of our method on the LAVIB dataset with different values of $K$ for the number of clusters employed during pretraining. The clustering performance is evaluated against ground-truth complexity labels extracted from 1280$\times$720 videos. It may be observed that our method only demonstrates a slight variation in performance with different values of $K$, indicating that our method is robust to the choice of $K$ during pretraining. The best performance is achieved with $K=10$, which is used as the default value for our method in all experiments. The table also shows the minimum, maximum, and median cluster sizes for each value of $K$ during pretraining. It may be observed that the median cluster size is lower than the mean cluster size for all values of $K$, indicating a right-skewed distribution of cluster sizes. Supplementary Material provides additional ablation studies on the effect of different input resolutions during downstream training and evaluation, and the effect of clustering algorithms to evaluate the clustering performance of the visual encoders.

%% file: sections/7_conclusion.tex
\section{Conclusion}
\label{sec:conclusion}
We have addressed the highly challenging problem of video encoding complexity clustering, which is crucial for optimizing video streaming and compression. We present Compression Echo Contrastive Learning (CECL), a novel self-supervised learning framework for pretraining video encoders to learn representations for video encoding complexity clustering. Our method leverages the responses of videos to compression, termed the `Compression Echo', as a supervisory signal to capture those spatio-temporal video characteristics most amenable to video compression. We introduced a dynamic pseudo-labeling strategy and a contrastive learning objective to effectively learn representations that capture encoding complexity characteristics. Through extensive experiments, we show that our method demonstrates improved performance over existing visual encoders and delivers strong bitrate and quality savings. We hope our method can serve as an initial step towards developing more efficient video encoding strategies.

\section{Limitations and Future Work}
Our proposed method is primarily designed for the problem of video encoding complexity clustering, and its applicability to other downstream tasks remains unexplored. While the method has shown promising results in our experiments, the clustering performance could be further improved. Additionally, our method relies on the availability of high-resolution videos and precomputation of compressed videos before pretraining, which may limit its scalability. In the future, we plan to investigate more precise and efficient pretraining strategies along with alternative supervisory signals that do not require compression. Furthermore, we aim to explore the generalizability of our method to other resolutions, presets, and video codecs.

%% file: sections/appendix.tex
\section{Changes to Bjontegaard Metric Calculation for Symmetry}
\label{sec:clustering_algorithm}
In the clustering algorithm proposed by researchers from Meta \cite{Towards-Perceptually-Optimized-Compression-Of-User-Generated-Content}, the authors employed BD metrics \cite{BD-Metric} as distance metrics to cluster videos together. It is important to note that the units for BD-VMAF and BD-Rate are different, with BD-VMAF typically being on the VMAF scale and BD-Rate being in percentage. This difference in units can lead to confusion when interpreting the results of the clustering algorithm, especially when comparing clusters or making decisions based on these metrics. Specifically, the BD-VMAF values between videos `i' and `j' are complementary (times -1) to the BD-VMAF values between videos 'j' and 'i'. This means that if BD-VMAF(i, j) = x, then BD-VMAF(j, i) = -x. However, this property does not hold for BD-Rate, as the percentage bitrate changes are not symmetric. This asymmetry can lead to confusion when interpreting the BD-Rate values, especially when comparing clusters or making decisions based on these metrics. Hence, we propose a modified BD-Rate calculation to ensure that symmetry is maintained. The algorithm \ref{Algorithm:BD-Rate} outlines the steps to compute the BD-Rate between two videos. In our experiments, we will use this modified BD-Rate calculation to ensure our calculations stay in an exponential scale, thereby avoiding percentages and maintaining the symmetry of the BD-Rate values between two videos. The rest of the clustering algorithm remains unchanged, and we will continue to use the same clustering techniques as proposed by Meta \cite{Towards-Perceptually-Optimized-Compression-Of-User-Generated-Content} in our experiments.
\begin{algorithm}[!ht]
\caption{Modified BD-Rate for Videos for Symmetry}\label{Algorithm:BD-Rate}
\begin{algorithmic}[1]
	\State \textbf{Input:} Videos $\text{V}_{i}$, $\text{V}_{j}$
	\State avgDiffPercent = BD-Rate($\text{V}_{i}$, $\text{V}_{j}$) \Comment{BD-Rate between videos $\text{V}_{i}$ and $\text{V}_{j}$}
	\State avgExpDiff = $\log(1 + \text{avgDiffPercent}/100)$ \Comment{Convert to exponential scale}
	\State \textbf{Output:} avgExpDiff
\end{algorithmic}
\end{algorithm}

\section{Dataset Preparation}
\label{sec:dataset_preparation}
\begin{figure*}[!ht]
	\centering
	\includegraphics[width=\textwidth]{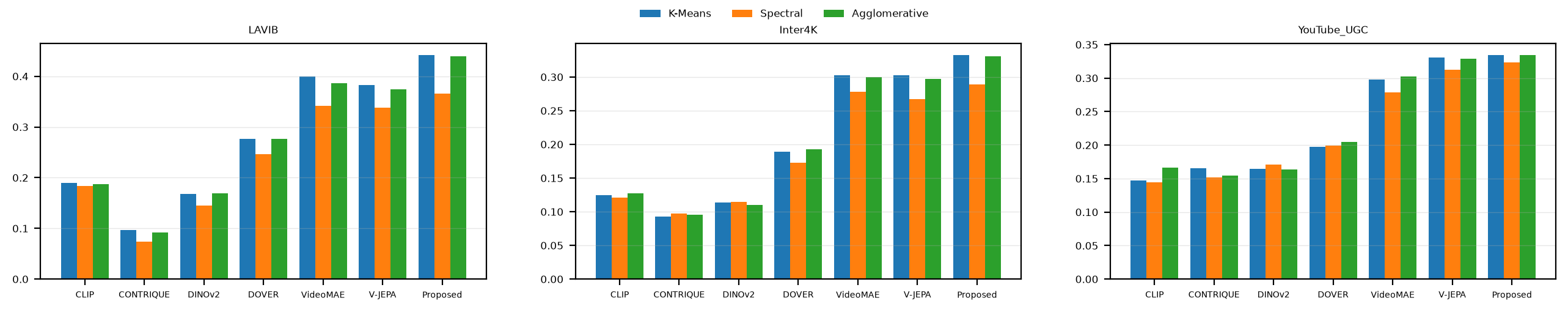}
	\caption{Performances of label prediction using various clustering algorithms using trained video representations against the ground truth cluster labels.}
	\label{fig:clustering_comparison}
\end{figure*}

\begin{figure*}[!ht]
	\centering
	\includegraphics[width=\textwidth]{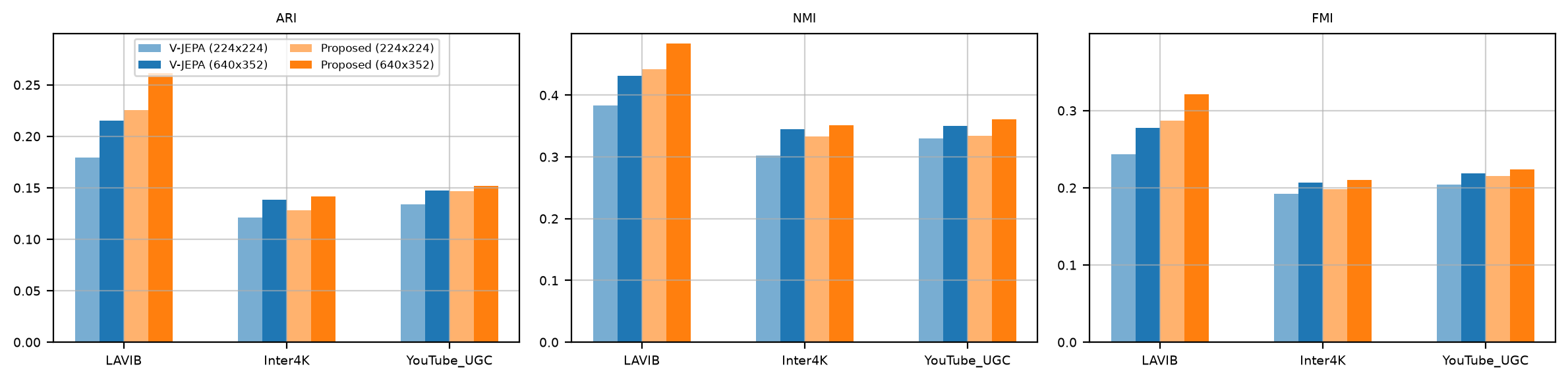}
	\caption{Impact of resolution on clustering performance on the LAVIB dataset.}
	\label{fig:resolution_impact}
\end{figure*}
In this section, we describe the dataset preprocessing procedures for various datasets used in our experiments. Since each dataset has its own unique characteristics, we applied different preprocessing steps to extract high-quality video scenes from each video in the dataset. The detailed preprocessing steps for each dataset are as follows:
\begin{itemize}
	\item \textbf{LAVIB} \cite{LAVIB} is a LArge-scale Video Interpolation Benchmark (LAVIB) for the low-level video task of Video Frame Interpolation (VFI). We randomly sampled 40,000 videos from the LAVIB dataset with a minimum resolution of 720p and a bitrate of at least 4 Mbps. We center-cropped the videos to a resolution of 1280$\times$720 from their original resolution of $1296\times1296$.
	\item \textbf{OpenVid} \cite{OpenVid} is a large-scale high-quality video dataset for text-to-video generation. We randomly sampled 40,000 videos from OpenVid having a minimum resolution of 1080p and a bitrate of at least 4 Mbps. We temporally cropped the videos to at most 60 frames and spatially resized them to a resolution of 1920$\times$1080.
	\item \textbf{Inter4K} \cite{Inter4K} dataset contains 1000 4K videos for video frame interpolation (VFI) and video super resolution (SR). Since each video may contain multiple scenes, we first divided the dataset into shots/scenes. We defined scenes as videos between two I-frames, including the I-frame at the beginning (left side). We only saved scenes having at least 60 frames. Among these, we only retrained those videos having bitrates no less than the 1st percentile of the bitrate distribution (as the dataset is small). We temporally cropped videos to at most 60 frames and spatially resized them to 1920x1080. In total, we obtained 1393 videos from the Inter4K dataset after preprocessing.
	\item \textbf{YouTube-UGC} \cite{YouTube-UGC} dataset contains User Generated Content sampled from YouTube. Similar to Inter4K, we first divided the dataset into shots or scenes, then retrained only those having bitrates no less than the 1st percentile of the bitrate distribution. We temporally cropped these videos to no more than 60 frames and spatially resized them to 1920$\times$1080. In total, we obtained 650 videos from the YouTube-UGC dataset after preprocessing.
\end{itemize}
We employed the FFmpeg library to conduct all video processing tasks, including cropping, resizing, and bitrate filtering. The resulting videos were used for model training and validation in our experiments.

\section{Training and Evaluation Settings}
\subsection{AttentivePooler Architecture}
Our AttentivePooler consists of a linear layer at the start to embed the video representations from the visual encoder to a lower dimension (256). This is followed by a simple CrossAttention block with 4 heads and an MLP ratio of 1.0. Following \cite{SimCLR}, we applied a projection head with 2 linear layers and a ReLU activation in between to project the video representations to a lower dimension (128) to compute the contrastive loss. During inference, we discarded the projection head and used the output of the CrossAttention block as the video representation for clustering.

\subsection{Optimization Parameters}
Due to space constraints in the main text, we detail the optimization hyperparameters used for both pretraining and downstream training in Table \ref{tab:optimization_hyperparameters}. We employed the AdamW optimizer \cite{AdamW} for both pretraining and downstream training, along with learning rate and weight decay schedulers. We also employed mixed precision training to reduce memory usage and speed up training.

\begin{table}[!ht]
	\centering
	\caption{Detailed optimization hyperparameters for Pretraining and Downstream Training.}
	\footnotesize
	\setlength{\tabcolsep}{3pt}
	\resizebox{\columnwidth}{!}{
	\begin{tabular}{lcc}
		\toprule
		\textbf{Hyperparameter} & \textbf{Pretraining} & \textbf{Downstream Training} \\
		\midrule
		Epochs & 100 & 10 \\
		$\tau$ (Temperature) & 0.07 & 0.07 \\
		Optimizer & AdamW & AdamW \\
		Base Learning Rate & 2e-4 & 2e-4 \\
		LR Scheduler & WarmupCosine & Cosine \\
		Warmup Epochs for LR Scheduler & 5 & - \\
		Layer-wise LR Decay & 0.75 & - \\
		Starting Learning Rate & 1e-5 & 1e-5 \\
		Final Learning Rate & 1e-5 & 1e-5 \\
		\midrule
		Base Weight Decay & 0.05 & 0.02 \\
		WD Scheduler & Cosine & Cosine \\
		Final Weight Decay & 0.5 & 0.05 \\
		\midrule
		Gradient Clipping & 10.0 & 1.0 \\
		Start EMA Decay & 0.998 & - \\
		End EMA Decay & 1.0 & - \\
		Mixed Precision Training & True & True \\
		\bottomrule
	\end{tabular}}
	\label{tab:optimization_hyperparameters}
\end{table}

\section{Additional Ablation Studies}
Figure \ref{fig:clustering_comparison} shows the clustering performance of various methods on 720p labels using different clustering algorithms during inference. It may be observed that k-Means and Agglomerative clustering algorithms demonstrate competitive performance, while Spectral clustering performs poorly. In our experiments, we employ k-means clustering for all methods during inference. 

Figure \ref{fig:resolution_impact} shows the impact of input resolution on clustering performance during attention-probing on the LAVIB dataset. It may be observed that the clustering performance of both V-JEPA \cite{V-JEPA} and our method improves with increasing input resolution, indicating that higher input resolutions provide more information to learn better global representations for clustering. However, it should be noted that higher input resolutions also increase the computational cost and memory requirements during training, resulting in a larger number of image patches/tokens to be processed by the visual encoder.